\documentstyle[prd,aps,preprint]{revtex}
\tightenlines
\newcommand {\be}{\begin{eqnarray}}
\newcommand {\ee}{\end{eqnarray}}
\input epsf.tex
\def\DESepsf(#1 width #2){\epsfxsize=#2 \epsfbox{#1}}
\begin{document}

\draft
\preprint{\vbox{
\hbox{UMD-PP-00-024}
\hbox{SMU-HEP-99-07}
}}
\title{ TeV Scale Quantum Gravity and Mirror Supernovae as Sources of
Gamma Ray Bursts}
\author{ Rabindra N. Mohapatra$^1$, Shmuel Nussinov$^2$ and Vigdor L.
Teplitz$^3$ }

\address{$^1$Department of
Physics, University of Maryland, College Park,
MD-20742\footnote{e-address:Rmohapat@physics.umd.edu}\\
$^2$ Department of Physics, Tel Aviv University, Tel Aviv, Israel \\
and Department of Physics, University of South Carolina, Columbia,
SC-29208\\
$^3$ Department of Physics, Southern Methodist University, Dallas,
TX-75275.}
\date{September, 1999}
\maketitle
\begin{abstract}
Mirror matter models have been suggested recently as an explanation of
neutrino puzzles and microlensing anomalies.  We show that  mirror
supernovae can
be a copious source of energetic gamma rays if one assumes that
the quantum gravity scale is in the TeV range. We show that under certain
assumptions plausible in the mirror models, the gamma energies could be
degraded to the 10 MeV range (and perhaps even further) so as to provide
an explanation of observed gamma ray bursts. This mechanism for the origin
of the gamma ray bursts has the advantage that it neatly avoids the
``baryon load problem''.

\end{abstract}

\section{Introduction}

The origin of the gamma ray bursts (GRBs) observed for over three decades
still remains unclear\cite{piran}. The GRBs are short,
intense photon bursts with photon energies in the keV and MeV range
although
bursts with energy spectra extending above a GeV have been observed.
The isotropy and $\frac{dN}{dV}$(intensity) distributions and the high
redshift galaxies associated with some GRBs indicate that the
sources of GRBs are located at cosmological distances.
The specific nature of the sources remains however unclear.

If unbeamed, the sources must emit total $\gamma$-ray energies 
of $10^{51}$ to $10^{53}$ ergs\cite{piran}\footnote{Beaming reduces this
by $\frac{\Delta \Omega}{4\pi}$ and increases the required burst rate by
$\frac{4\pi}{\Delta \Omega}$ over the few per day seen in the universe.} 
This is very much reminiscent of typical
supernova energies. However, most supernovae (e.g. type II supernovae) 
cannot be these sources,
since $\gamma$-rays with typical radiation lengths of 100 gm/cm$^2$ cannot
penetrate the large amount ($\sim 10~M_{\odot}$) of overlying ejecta.

Many of the models for unbeamed (beamed) GRBs use massive compact sources
to produce neutrinos which annihilate to form fireballs of $e^+e^-$'s and
$\gamma$'s\cite{goodman,eichler}. The fireballs expand and cool
adiabatically, until the temperature (or the transverse energy) is low
enough so that the $e^+e^-$ annihilate into the $\gamma$'s. To avoid
the `baryon load'' problem and the absorption of $\gamma$'s, fairly ``bare
collapses'' are required\cite{dar}. Accretion induced collapses and binary
neutron star mergers\cite{piran} were considered but it is not clear
whether these are sufficiently ``baryon clean''.  One ``baryon clean''
source candidate based novel particle physics is a neutron star to strange
quark star transition.

Other recent suggestions\cite{kluz,bli} invoked the existence of sterile
neutrinos\cite{sterile}. If the emitted neutrinos undergo maximal
oscillation to the sterile neutrinos\cite{kluz}, the latter can penetrate
the baryon barrier and subsequently normal neutrinos will appear via the
$\nu_s-\nu$ oscillation. In this scenario, the last ``back'' conversion
occur at relatively large distances \footnote{Both $\nu\rightarrow \nu_s$
and $\nu_s\rightarrow \nu$ are quenched by dense matter if $\Delta m^2\leq
10^4$ eV$^2$\cite{yv}.} and the $\nu\bar{\nu}\rightarrow e^+e^-$ which
goes like $R^{-8}$\cite{goodman} is inefficient\footnote{Disklike (and
beamed) geometry may partially alleviate this problem.}. Similar
difficulties are encountered by models utilizing exact ``mirror''
symmetric theories\cite{foot} where the sterile (mirror) neutrinos 
emitted in a mirror star collapse oscillate into ordinary neutrinos.

In this note, we propose another GRB scenario in the context of the
asymmetric mirror models\cite{berezhiani}. It utilizes the
conversion of $\nu'+\bar{\nu}',\gamma'+\gamma'\rightarrow e^+e^-,
\gamma\gamma$ etc inside the mirror star, where primed symbols denote
mirror particles. Since the familiar electrons and photons do
not interact with mirror matter, the expanding fireball is not impeded
and we have an ideal bare collapse. The resulting photons expected to have
initial energies of $\approx$ GeV, can be processed in this expansion down
to the MeV part of the GRB spectra observed. Furthermore, if the source
is embedded in the disk of a galaxy, further degrading can take place due
to the ``minibaryon load'' of the disk resulting in keV gamma rays as well
as possibly structure in the gamma ray spectrum.

The key requirement is that the conversion process be fast enough so that
a finite fraction of the collapse energy is indeed converted into ordinary
matter. As we will see this naturally obtains\cite{sila} if we can have a
low scale (of order of a TeV) for quantum gravity\cite{nima}\footnote{
In the p-Brane construction, ordinary and mirror matter could reside on 
two sets of branes\cite{moha} with a relatively large (compared to
$\Lambda^{-1}\approx ~(TeV)^{-1}$) separation $r_0$. The gauge group is
of the form $G\rightarrow G_{matter}\times G_{mirror}$ where each $G=
SU(3)\times SU(2)_L\times U(1)_Y$. The detailed model implementing this
scenario will have to be such that it can lead to enhanced amplitude for
the four Fermi operators that lead to familiar particle production via the
collision of mirror particles whereas suppressed coefficient for the ones
that lead to neutrino mixing. The latter in general involve exchange of
fermions and the desired suppression is therefore not implausible. We
thank Markus Luty for discussions on this point.} 

In section 2 we give a brief review of the assumptions of mirror matter
models within which we work. In section 3 we outline our scenario,
computing the initial $\gamma$ energies, and a brief discussion of
possible fireball mechanism for degradation of the photon energies. We
also discuss the effect of a baryon cloud (``mini-baryon load'') which can
lead to further degradation of gamma energies. We work within the
framework of TeV scale gravity using the results of Silagadze\cite{sila}
for the production of familiar matter from mirror matter. We conclude in
section 4 with a brief discussion.

\section{Asymmetric mirror model and large scale structure in the mirror
sector}

Let us begin with a brief overview of the asymmetric  mirror matter
model and the  the parameters describing fundamental forces in the mirror
sector. In asymmetric mirror matter models\cite{berezhiani}, one
considers a duplicate version of the standard model
with an exact mirror symmetry\cite{lew} which is broken\cite{berezhiani}
in the process of gauge symmetry breaking. Denoting
all particles and parameters of the mirror sector by
a prime over the corresponding familiar sector symbol (e.g. mirror quarks
are $u',d',s',$ etc and mirror Higgs field as $H'$, mirror QCD scale as 
$\Lambda'$) we assume that $<H'>/<H>=\Lambda'/\Lambda\equiv
\zeta$\cite{tep2}. This is admittedly a strong assumption for which there
is no
particle physics proof, but it does provide a certain degree of economy.
Of course, if one envisioned the weak interaction symmetry to be broken
by a new strong interaction such as technicolor in both sectors,
then it is possible to argue that such a relation emerges under
certain assumptions.

There also exists a cosmological motivation for assuming
$<H'>/<H>=\Lambda'/\Lambda\simeq 15$. One can show that in this case
the mirror baryons can play the role of cold dark matter of the
universe\cite{tep2,bere}.
The argument goes as follows: one way to reconcile the mirror universe
picture with the constraints of big bang nucleosynthesis (BBN) is to
assume  asymmetric inflation with the reheating temperature in the mirror
sector being slightly lower than that in the normal one\cite{reheat}. 
Taking the allowed extra number of neutrinos at the BBN to be 1 implies
$(T'_R/T_R)^3\leq 0.25$. One can then calculate the contribution of the
mirror baryons to $\Omega$ to be
\begin{eqnarray}
\Omega_{B'}\simeq (T'_R/T_R)^3 \zeta \Omega_B
\end{eqnarray}
Since one expects, under the above assumption, the masses of the proton
and neutron to scale as the
$\Lambda$ in both sectors, if we assume that $\Omega_B\simeq 0.07$, then
this implies $\Omega_{B'}\simeq 0.26$ leading to a total matter content
$\Omega_m\simeq 0.33$. Thus familiar and mirror baryons together could
explain the total matter content of the universe without need for any
other kind of new particles.

An important implication of this class of mirror models is that the
interaction strengths of weak as well as electromagnetic processes
(such as Compton scattering cross sections etc) are much smaller than that
in the familiar sector. This has implications for the formation of
structure in the mirror sector.

Structure formation in a similar asymmetric mirror model was studied
in Ref.\cite{tep1} where it was  shown that despite the weakness
in the mirror particle processes, there are cooling mechanisms that allow
mirror condensates to form as the universe evolves. The basic idea is that
the mirror matter provides gravitational wells into which the familiar
matter gets attracted to provides galaxies and their clusters. However due
to weakness of the physical processes, the mirror matter is not as
strongly dissipative as normal matter. So for instance in our galaxy, the
familiar matter is in the form of a disk due to dissipative processes
whereas mirror stars which form the halo are not in disk
form. In contrast, in the symmetric mirror model\cite{foot}, the mirror
matter would also be in a disk form and therefore could not help in
explaining
observed spherical galactic halos. Furthermore, since mirror matter
condensed first in view of the lower temperature, it is reasonable to
expect that mirror star formation largely
took place fairly early (say $z\geq 1$) and the subsequent rate is much
lower. In what follows to understand the observed GRBs we would require
a mirror star formation rate of about one per million year per galaxy (to
be contrasted with about 10/year/galaxy for familiar stars). 

In the asymmetric mirror model, it has been shown that there are simple
scaling laws (first reference in \cite{tep2}) for the parameters of the
mirror stars: (i) the mass of the mirror stars scale as $\zeta^{-2}$; (ii)
the radius of the mirror stars also
scales like $\zeta^{-2}$ whereas (iii) the core temperature scales
slightly faster than
$\zeta$. Here $\zeta$ denotes the ratio of the mass scales in the
mirror and familiar sectors and is expected to be of order 15-20 from
considerations of neutrino physics\cite{berezhiani}. Due to the higher
temperature of the mirror stars, they will ``burn''
much faster and will reach the final stage of the stellar evolution
sooner. Because of the $\zeta^{-4}$ decrease of weak cross sections
and the increase in particle masses we do not expect mirror star collapse
to result in explosion.  Rather there should be neutrino emission and
black hole formation.  Thus we would expect that there will be an abundant
supply of mirror ``supernovae.'' We will show in the next section that
these could be the sources of the GRBs.

\section{Low quantum gravity scale and production of familiar photons in
mirror supernovae}

In a mirror supernova, one would expect most of the gravitational binding
energy to be released via the emission of mirror neutrinos as in the
familiar case. However, in the asymmetric mirror matter model, we expect
the temperature
of the collapsing star to be higher.  We have $NT=GM^2/R$ where $N$ is
the number of mirror baryons in the star (about $M_{\odot}/\zeta m_p$).
At $\zeta = 10$ the maximum mirror star mass is about $M_{\odot}$ so that
$T$ is about a $GeV$ where we have taken the radius of the collapsed
mirror star to be about a kilometer. Let us now estimate the
production cross section for the familiar photons in the collision of the
mirror photons in the core.

The most favorable case occurs if we assume that the quantum gravity scale
is in the TeV range\cite{sila}. In this case, assuming two extra
dimensions\cite{nima} and following reference \cite{han}, we estimate
the cross section
$\sigma_{{\gamma'}{\gamma'}\rightarrow \gamma\gamma}$ to be,
\begin{eqnarray}
\sigma_{{\gamma'}{\gamma'}\rightarrow \gamma\gamma}
\simeq \frac{1}{10}\frac{s^3}{\Lambda^8}
\end{eqnarray}
where $s$ is the square of the total center of mass energy. For $s= 1$
GeV$^2$ and $\Lambda \simeq 1$ TeV, we get,
$\sigma_{{\gamma'}{\gamma'}\rightarrow \gamma\gamma}\simeq
10^{-52}$ cm$^2$. We estimate the rate of energy loss per unit volume to
into familiar, not mirror, photons to be roughly
\begin{eqnarray}
\frac{dQ}{dtdV}\simeq
cn^2_{\gamma'}2E_{{\gamma'}}\sigma_{{\gamma'}{\gamma'}
\rightarrow\gamma\gamma}
\end{eqnarray}
Multiplying by the volume of the one kilometer black hole gives about 
 $10^{52}$ erg/s. This energy is of
the right order  of magnitude for the total energy release in the
case of unbeamed or mildly beamed GRBs. However the initial energy of 
individual photons obtained via $\nu'\rightarrow \gamma$ conversion is
essentially that of the mirror neutrinos i.e. $E_{\gamma}(t=0)\approx
E_{\nu'}\approx 3 T_{mirror}$. The spectrum of the latter- just like that
of ordinary neutrinos obtained in the core cooling of ordinary type II
supernovae- is expected to be roughly thermal with $T_{mirror}\approx $
GeV, which is roughly 100 times higher than for familiar collapse.

While in some GRBs, photons of energies in the range of GeVs to TeVs have
been observed, the bulk of the spectrum is in the MeV/keV region.
Reprocessing
the initial photons leading to energy degradation is therefore important.
Two distinct mechanisms contribute to reprocessing: (i) Fireball evolution
and (ii) Overlying putative familiar material. Let us discuss both these
mechanisms.

\bigskip
\noindent{\it Mechanism (i):}

\bigskip

At t=0, we have, because of universality of gravitational interactions
an equal number of familiar $e^+e^-$ produced with the photons. The
resulting dense $e^+e^-\gamma$ ``fireball'' constitutes a highly
opaque plasma. There is an extensive literature dealing with the evolution
of such fireballs\cite{goodman,dar}. In the case where this evolution is
free from the effects of overlaying matter (i.e. the effects of (ii) are
negligible), the discussion becomes almost model independent and many
features can be deduced from overall energetics and thermodynamic
considerations. Thus at t=0 when a fraction $\epsilon$ of the mirror
neutrinos convert to $\gamma$'s (and/or $e^+e^-$'s), the latter have a
blackbody spectrum with temperature $T_{\nu'}$. However the overall
normalization,  i.e. the energy density 
\begin{eqnarray}
U_{\gamma}= \epsilon U_{\nu'} = \epsilon a T^4_{\nu'}
\end{eqnarray}
falls short  by a factor $\epsilon$ of the universal black body energy
density at such a temperature. Fast processes of the form
$\gamma\gamma\rightarrow e^+e^-\rightarrow 3\gamma$ (allowed in the
thermal environment) will then immediately reequilibrate the system at
\begin{eqnarray}
T_{\gamma} \sim \epsilon^{\frac{1}{4}} T_{\nu'} \approx
\left(\frac{1}{3}-\frac{1}{30}\right) T_{\nu'}
\end{eqnarray}
(corresponding to GRB energies between $10^{48}-10^{52}$ ergs and mirror
supernova energies of $10^{52}-10^{53}$ ergs).
Subsequent evolution can further increase $N_{\gamma}$ and correspondingly
decrease $\overline{E}_{\gamma}$ down towards the MeV range. Independent
of this, mere expansion reduces the transverse photon energy according
to $E^{tr}_{\gamma}\approx (R/r)T_{\gamma}(t=0)$, where R is the size of
the source and $r$ is the current $\gamma$ location. (The last expression
which parallels that for adiabatic cooling simply reflects the geometrical
convergence of trajectories of colliding $\gamma$'s which become more and
more parallel with distance $r$.) Since $E_{tr}$ controls the center of
mass energy of the $\gamma \nu$ collisions, the $\gamma\gamma\rightarrow
e^+e^-$ processes become kinematically forbidden and the density of
$e^+e^-$ pairs
falls exponentially i.e. $n_{e^+e^-}\approx e^{-\frac{m_e}{T_{tr}(r)}}$
eventually leaving freely propagating $\gamma$'s.

\bigskip
\noindent{\it Mechanism (ii):}
\bigskip

A ``mini-baryon load'' of familiar material encountered by the outgoing
$\gamma$'s could further reduce the photon energy. Also the presence such
matter in conjunction with mild beaming could induce the very short time
structure often observed.

 In order to have an effective degrading of the emitted photon energies,
we will need an appropriate density of familiar matter which can be
estimated as follows. Let us assume a density profile of the form:
\begin{eqnarray}
\rho(R)=\frac{\rho_0 R^2_0}{R^2+R^2_0}
\end{eqnarray}
Then we demand the constraint that $\int \rho(R) dR \simeq 100$ gm/
cm$^2$ where 100 gm/cm$^2$ represents the radiation length of photons in
matter. This implies $\rho_0 R_0 \simeq 100$ gm/cm$^2$. 
The kinematical requirement of having comoving baryonic plus fireball
system requires
\begin{eqnarray}
\gamma_B\equiv \frac{f W_{GRB}}{M_{Baryo}}\approx \gamma_{Fireball}\approx
\frac{E_{e^+e^-}}{2m_e}
\end{eqnarray}
where $f$ is the fraction of energy imparted to baryons and $\gamma_B$ is 
the Lorentz factor. Using
$M_{baryo}\approx \frac{4\pi}{3}(\rho_0R_0)R^2_0$, we find
\begin{eqnarray}
R_0=10^{12}~cm~\left[\frac{(W/(10^{50}~ergs))}{(E/100~MeV)
(\rho_0R_0/100~gm~cm^{-3})}\right]^{1/2}
\end{eqnarray}
so that for the nominal values of the total GRB energy, the fireball
processed energy of individual $e^+, e^-,\gamma$ and the column density,
we find $R_0=10^{12}$ cm so that $\rho_0= 10^{-10}$ gr/cm$^{3}$ and
$M_{baryo}\approx 10^{25} ~gr \simeq 10^{-8} M_{\odot}$.
 It is interesting to note that in the present scenario, GRB's originating
from mirror supernovae in the galactic halos, which most likely would not
face the ``minibaryon load'', may have only the first stage i.e. energy
degradation by fireball mechanism and hence will have a harder spectra and
smoother time profile. (Clearly discerning such a component in the 
GRB population will be quite interesting.) On the other hand the GRBs
originating
from supernovae in the disk of galaxies will have degradation due to both
mechanisms and therefore more structure in the spectra as well as a softer
spectra.

Beyond the immediate neighbourhood of the mirror star
there would be further energy degradation from interaction with
interstellar matter ranging from molecular clouds to interstellar comets.
There is not however sufficient material in one kilopersec to overcome the
small value of the Thompson cross section i.e. $n_e\sigma_T \ell\sim
10^{-2}$ as against a required value of one.

\section{Discussion}

Section 3 shows, we believe, that mirror matter supernovae, within the
asymmetric mirror matter model, can provide a plausible explanation for
gamma ray bursts.  The scenario requires some coupling between the mirror
and familiar sectors.  In Section 3, we have used the couplings provided
by TeV
range quantum gravity following the estimate of reference (\cite{sila}),
but other coupling mechanisms (such as a small $\gamma-\gamma'$ mixing) 
might be possible as well. Given TeV scale
gravity, it is noteworthy that the same value of $\zeta$ required by other
"manifestations" of mirror matter gives both an appropriate upper limit to
the energy of the familiar gammas produced and an appropriate cross
section section for their production.  A major advantage of this GRB
explanation is that it solves the baryon load problem in a natural way.
In this model, we would expect production of GeV neutrinos at nearly the
same rate as $e^+e^-$ and $\gamma\gamma$ etc. For GRBs located in our
galaxy, they should be observable in detectors such as Super-Kamiokande.

If this model is correct, given the short lifetime of the mirror
stars\cite{tep2}, the GRB frequency of $10^{-6}$/year/galaxy must be a
result of low mirror star formation rate, which as mentioned above is not
an unreasonable assumption.

Finally, it is tempting to speculate that, if the primary GRB mechanism is
to produce a fireball in the many MeV temperature range, there should
exist a GRB population with temperatures in that range. In view of the
fact that most of the data on GRBs comes from BATSE detector which
triggers mostly on $\gamma$'s below 300 keV, it appears that such a
population is not necessarily excluded by current data.

This possibility that mirror matter can explain GRBs adds to a growing
list of arguments that asymmetric mirror matter should be taken seriously.
These include: (1) the requirement in many string theories
that mirror matter exist; (2) the fact that the same range for $\zeta$
that was required in Section 3 for GRBs gives a mirror neutrino at the
proper mass difference from $\nu_e$ to be the sterile neutrino responsible
for simultaneously solving all the neutrino puzzles; (3) the fact that the
same range of $\zeta$ gives an appropriate amount of dark matter to give
an overall $\Omega_M$ in the range 0.2 to 0.3; and (4) the fact that the
same range of $\zeta$ gives an explanation of the MACHO microlensing
events as being caused by mirror black holes of about $M_{odot}/2$ mass.

\bigskip
\noindent {\it Acknowledgments}

We appreciate a helpful communication from Tom Siegfried. 
The work of RNM is supported by a grant from the National
Science Foundation under grant number PHY-9802551 and the work of V. L. T.
is supported by the DOE under grant no. DE-FG03-95ER40908.

\end{document}